# An overview to Software Architecture in Intrusion Detection System

* Mehdi Bahrami[1], Mohammad Bahrami[2]

Department of Computer Engineering, I.A.U., Booshehr Branch, Iran

Bahrami[1];Shayan[2]@LianPro.com

**Abstract.** Today by growing network systems, security is a key feature of each network infrastructure. Network Intrusion Detection Systems (IDS) provide defense model for all security threats which are harmful to any network. The IDS could detect and block attack-related network traffic. The network control is a complex model. Implementation of an IDS could make delay in the network. Several software-based network intrusion detection systems are developed. However, the model has a problem with high speed traffic. This paper reviews of many type of software architecture in intrusion detection systems and describes the design and implementation of a high-performance network intrusion detection system that combines the use of software-based network intrusion detection sensors and a network processor board. The network processor which is a hardware-based model could acts as a customized load balancing splitter. This model cooperates with a set of modified content-based network intrusion detection sensors rather than IDS in processing network traffic and controls the high-speed.

**Keywords**: Intrusion Detection systems, Software Architecture, IDS, Network

## 1. Introduction

In this decade the Internet has experienced an explosive growth. Growing the network adds new services through Internet. These new services, add impact of security attacks [2]. Several algorithms and software systems are developed to protect network and systems from all attacks [4]. These services are provided through operating systems, software applications, such as: Antivirus, Firewall, Anti Spams and etc.

Intrusion Detection Systems (IDS) [3] are provided from around two decades ago, starting in 1980 with the publication of John Anderson's Computer Security Threat Monitoring and Surveillance [1]. Another paper, "An Intrusion Detection Model," published in 1987 which was provided a methodological framework and was useful for commercial products [1].

Several commercial investments are emerges for developing IDS such as [5, 6]. However, technology was limited to provide a cost-effective model. In this decade, several algorithms [7, 8] for encryption, decryption, public key exchange, secure socket layer, digital signature are developed. These algorithms help to improve IDS technology as well.

In [1], John McHugh et. al. compared an attack sophistication versus intruder technical knowledge which is shown in the figure 1.

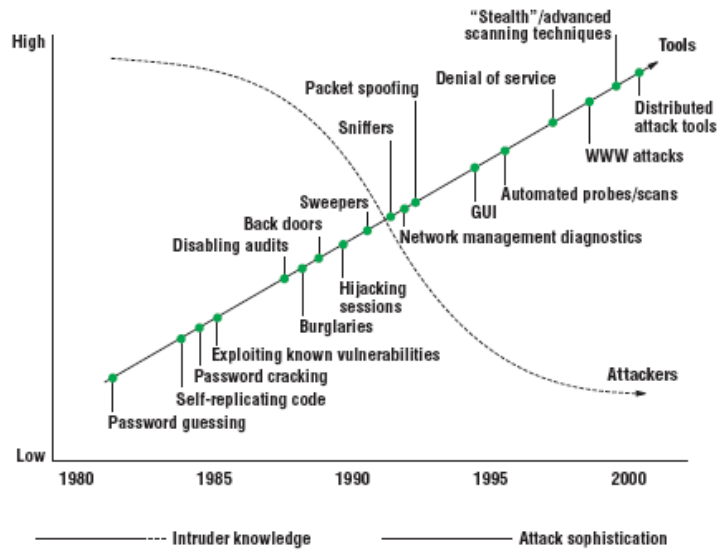

Figure 1. Attack sophistication versus intruder technical knowledge [1]

This chart, shows the complex model emerges through years and shows the simple model could not maintenance safety for a system.

To provide a safe model we need complex model which could use knowledge based model, such as data mining on the network protocols. For example, Lee et. al. in [2] provide a framework for constructing features and models for IDS. This model is shown in figure 2.

The model provides a data mining audit data framework for automated models in intrusion detection systems. Data mining allows the IDS to discover threat on the network.

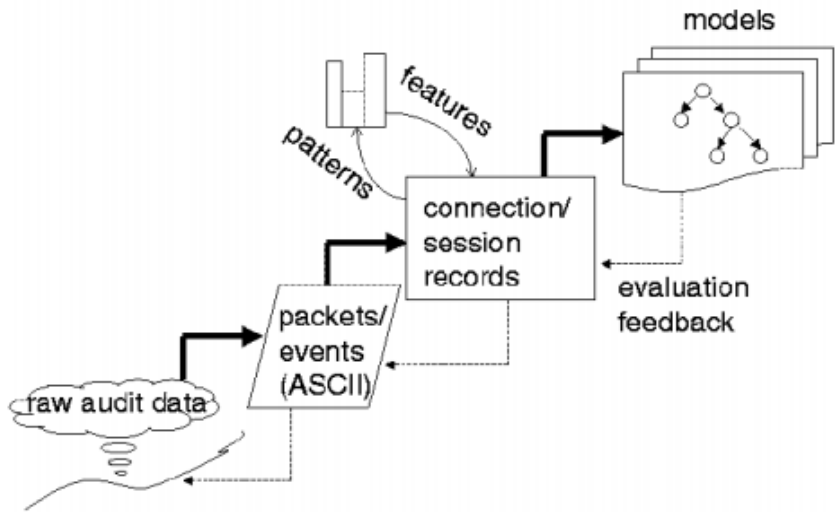

Figure 2. MADAM ID [2]

IDS could provide through software-based or hardware-based. The hardware-based could be provided through network connection and control over the network. The important part of network IDS or NIDS is far away from Operating Systems and other software applications. This segmentation allows NIDS works on without any conflict with software layer in a computer device.

## 2. Software Architecture

Software architecture provides a view of software components, connection between each component[26] and high-level design of software applications. However, different definitions of software architecture available [12-24] or for distributed system in [25], the following are a few of the most cited definition of software architecture:

• Bass, Clements, and Kazman, 1998: The software architecture of a program or computing system is the structure or structures of the system, which comprise software components, the externally visible properties of those components, and the relationships among them. By "externally visible" properties, we are referring to those assumptions other components can make of a component, such as its provided services, performance characteristics, fault handling, shared resource usage, and so on[9].

• Garlan and Perry, 1995: The structure of the components of a program/system, their interrelationships, and principles and guidelines governing their design and evolution over time [10].

• Garlan and Shaw, 1993: ...beyond the algorithms and data structures of the computation [11];

Simple software architecture that shows important components of an IDS is shown in the figure 2.

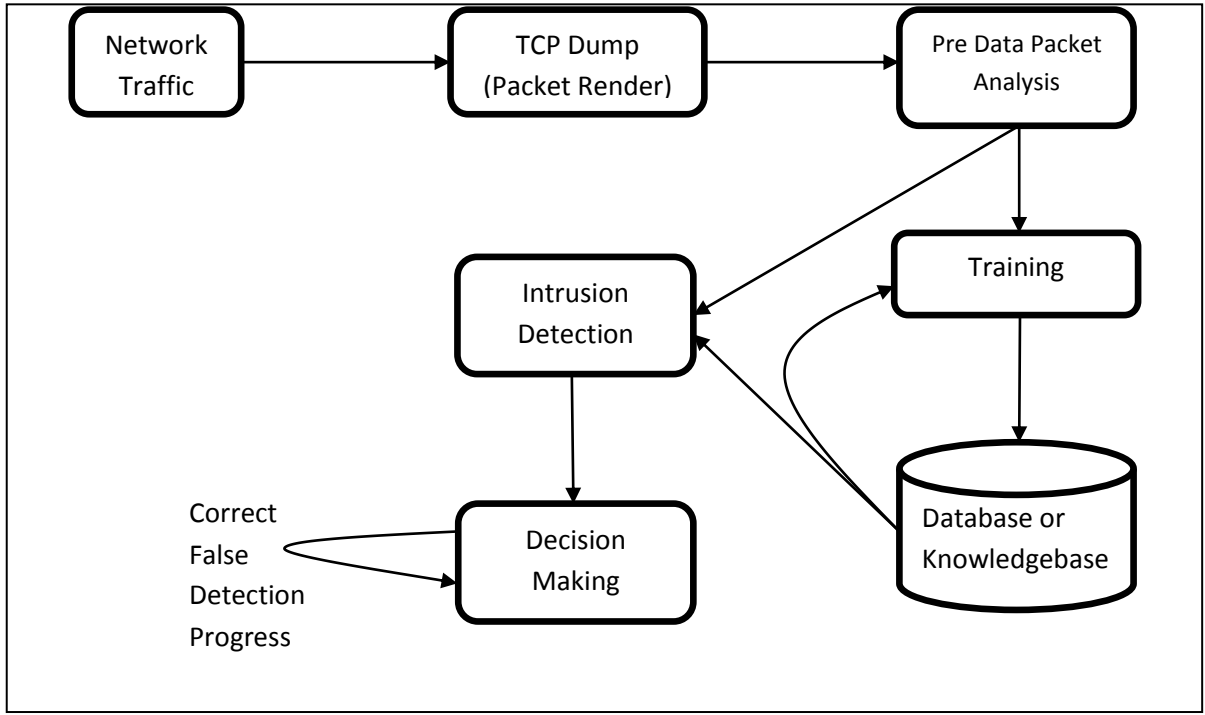

Figure 3. IDS software architecture overview

## 3. Cluster software architecture for IDS

Some other IDS model emerges for using IDS on the cluster of networks. The Cluster Head Module (CHM) is other model proposed and published in [31]. Some other model such as [27- 30] are emerged for control in a distributed models.

In this section we will review this architecture which is based on [31]. The CHM runs on each cluster-head node, and is responsible for the management of the cluster-member nodes in the cluster. CHM is also responsible for initiating cooperative intrusion detection and response action upon receiving a request from a cluster-member node.

The CHM is divided into six modules:

    I.    Cluster management module
   II.     Network information module
 III.    Mobile agent management module
 IV.    Global intrusion information module
  V.    Collaborative intrusion detection module
 VI.     Global intrusion response module. The interrelationships among the modules are depicted in Figure 4.

   The modules interact with each other via suitable messages in order to collect, store, process, and analyze the data. The functionalities of these modules are described below.

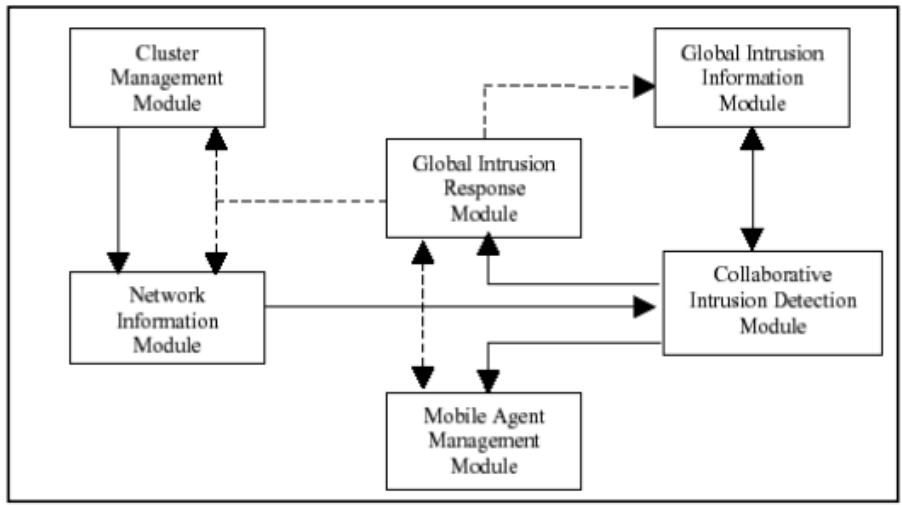

Figure 4. The architecture of the cluster head module [31]

The cluster management module manages the cluster by performing the functions such as [31]:

    I.   Registration of newly joined nodes
   II.   Supervision of elections in the cluster
  III.   Communication with other nodes in the cluster for cooperative intrusion detection. This module consists of three sub-modules:
        a.   Cluster-member registration sub module
        b.   Cluster-head election sub-module
        c.   Cluster member communication sub-module. In the rest of this subsection, the functionalities of these modules are described in the rest of this Section.

## 4. Conclusion

In this paper, we have presented a review of some important software architecture for intrusion detection systems as simple model, knowledge-based and cluster-based intrusion detection system architectures. The clustering of the network nodes makes message communication efficient. An efficient model allows to network working without any delay. Local detection also allows for detection of attacks for localized and could also work as global model with collaboration among the nodes in different clusters.

Citation:

Mehdi Bahrami, Mohammad Bahrami, "An overview to Software Architecture in Intrusion Detection System", International Journal of Soft Computing and Software Engineering [JSCSE], Vol. 1, No. 1, pp. 1-8, 2011